\documentclass[twocolumn,showpacs,amsmath,amssymb,pra]{revtex4}
\usepackage{amsmath}
\usepackage{graphicx}
\usepackage{color}
\usepackage{bm}
\usepackage[colorlinks,citecolor=blue]{hyperref}
\begin{document}
\title{Quantum memory using a hybrid circuit with flux qubits and NV centers}
\author{Xin-You L\"{u}$^{1}$}
\author{Ze-Liang Xiang$^{3,1}$}
\author{Wei Cui$^{1}$}
\author{J. Q. You$^{2,3,1}$}
\author{Franco Nori$^{1,4}$}
\address{$^1$CEMS, RIKEN, Saitama, 351-0198, Japan\;\;\;\;\;\;\;\;\;\;
\\
$^2$Beijing Computational Science Research Center, Beijing 100084, China
\\$^3$Department of Physics, State Key Laboratory of Surface Physics, Fudan University, Shanghai 200433, China
\\$^4$Physics Department, The University of Michigan, Ann Arbor, Michigan 48109-1040, USA}

\date{\today}

\begin{abstract}
We propose how to realize high-fidelity quantum storage using a hybrid quantum architecture including two coupled flux
qubits and a nitrogen-vacancy center ensemble (NVE). One of the flux qubits is considered as the quantum computing processor and the NVE
serves as the quantum memory. By separating the computing and memory units, the influence of the quantum computing process on the quantum memory can be effectively eliminated, and hence the quantum storage of an arbitrary quantum state of the computing qubit could be achieved with high fidelity. Furthermore the present proposal is robust with respect to fluctuations of the system parameters, and it is experimentally feasibile with currently available technology.
\end{abstract}


\maketitle
\section{Introduction}
Hybrid quantum circuits combining the advantages of atoms, spins, and solid-state devices could have applications for quantum
information processing and quantum computation \cite{0.5,1}. Examples of potential quantum memories in hybrid quantum circuits include the following: ultracold $^{87}$Rb atomic ensembles \cite{2}, polar molecular
ensembles \cite{3,3.5}, and spin ensembles \cite{4} with long coherence times. Examples of quantum computing processors performing quantum-gate operations typically involve superconducting qubits \cite{1,5.1,6,6.5,6.6} that couple strongly to electromagnetic fields. To couple the memory and computing units, previous proposals normally considered a common transmission line resonator as the quantum data bus and employed either electric-dipole or magnetic-dipole interactions \cite{1,4,7,7.1,7.5,8,8.1}.

Magnetic interactions are more desirable due to the sufficiently long coherence times achieved in systems with spin states storing quantum information. For example, nitrogen-vacancy (NV) centers in diamond have long electronic
spin lifetimes, narrow-band optical transitions, as well as the possibility of coherent manipulation at room temperature \cite{8.5,8.6,8.7,8.8}. However, magnetic interactions are inherently weaker compared with electric interactions, even though a strong magnetic coupling to ensembles of
spins has been achieved \cite{9}. 

Recently, some novel hybrid systems consisting of a SC flux qubit magnetically coupled to a nitrogen-vacancy center ensemble (NVE) were proposed \cite{11,12,12.1,12.2} and one of these implemented experimentally \cite{12.5} in order to enhance the corresponding magnetic-dipole interactions. Calculations in Ref.\,\cite{11} suggest that the magnetic coupling between a superconducting flux qubit and a single NV center could be about three orders of magnitude stronger than that associated with stripline resonators, thereby making the (flux qubit and NV center) system a possible building block for implementing quantum storage. However, in the single flux-qubit-NVE system in Ref.~\cite{11} [schematically shown in Fig.\,\ref{fig1}{\color{red}(a)}, with their energy levels in \ref{fig1}{\color{red}(b)}], the quantum ``computing'' processor and memory unit overlap, and hence it is difficult to individually
perform quantum computation without influencing the quantum memory. Moreover, the scalability of this quantum CPU$+$memory unit is problematic, because this would require coupling many quantum CPUs and memory units, respectively. 

Inspired by the above points, in this paper we propose an alternative method for realizing high-fidelity quantum storage by {\it separating} the quantum computing and memory units [see Fig.\,\ref{fig1}{\color{red} (c)}]. Here, the NVE acts as a spin-based quantum memory coupled to a flux qubit (qubit M) through a strong magnetic-dipole interaction; another flux qubit (qubit C) is the quantum computing processor coupled to the qubit M through the tunable coupling $J_{t}$. Notice that although the direct coupling between a coherent SC qubit and the atomic memory is still challenging, the physics of coupled flux-NVE system has been well established, theoretically and experimentally \cite{12.5}.

The major advantages of our proposal are as follows: (1) The quantum information is transferred between a SC flux qubit and a NVE, which is different from transferring quantum information from one SC qubit to another qubit as in recent experiments \cite{12.6}. 
(2) The quantum
computing and memory units are {\it separated} from each other, and thus the influence of the quantum computing process on the quantum memory can be
either drastically reduced or effectively eliminated. High-fidelity quantum storage can be realized without needing any additional operations on the quantum computing or memory units. (3) A large-scale quantum memory device is feasible by adding up or integrating individually the computing and memory units. (4) The present proposal is robust with respect to variations of some experimental parameters.

Even though this work focuses on a specific system, we would like to emphasize a more general message: a modular approach involving a qubit, coupler, and memory presents advances to the usual qubit-memory approach in hybrid systems.

The remainder of this paper is organized as follows. In section II we introduce the model under consideration and derive its effective Hamiltonian. In sections III and IV we discuss the realization of high-fidelity quantum storage based on resonant and dispersive interactions. Subsequently, the experimental feasibility of our proposal is discussed in section V. Finally, we conclude with a brief summary in section VI.

\section{Model and Hamiltonian}
\begin{figure}
\centering
\includegraphics[width=0.46\textwidth]{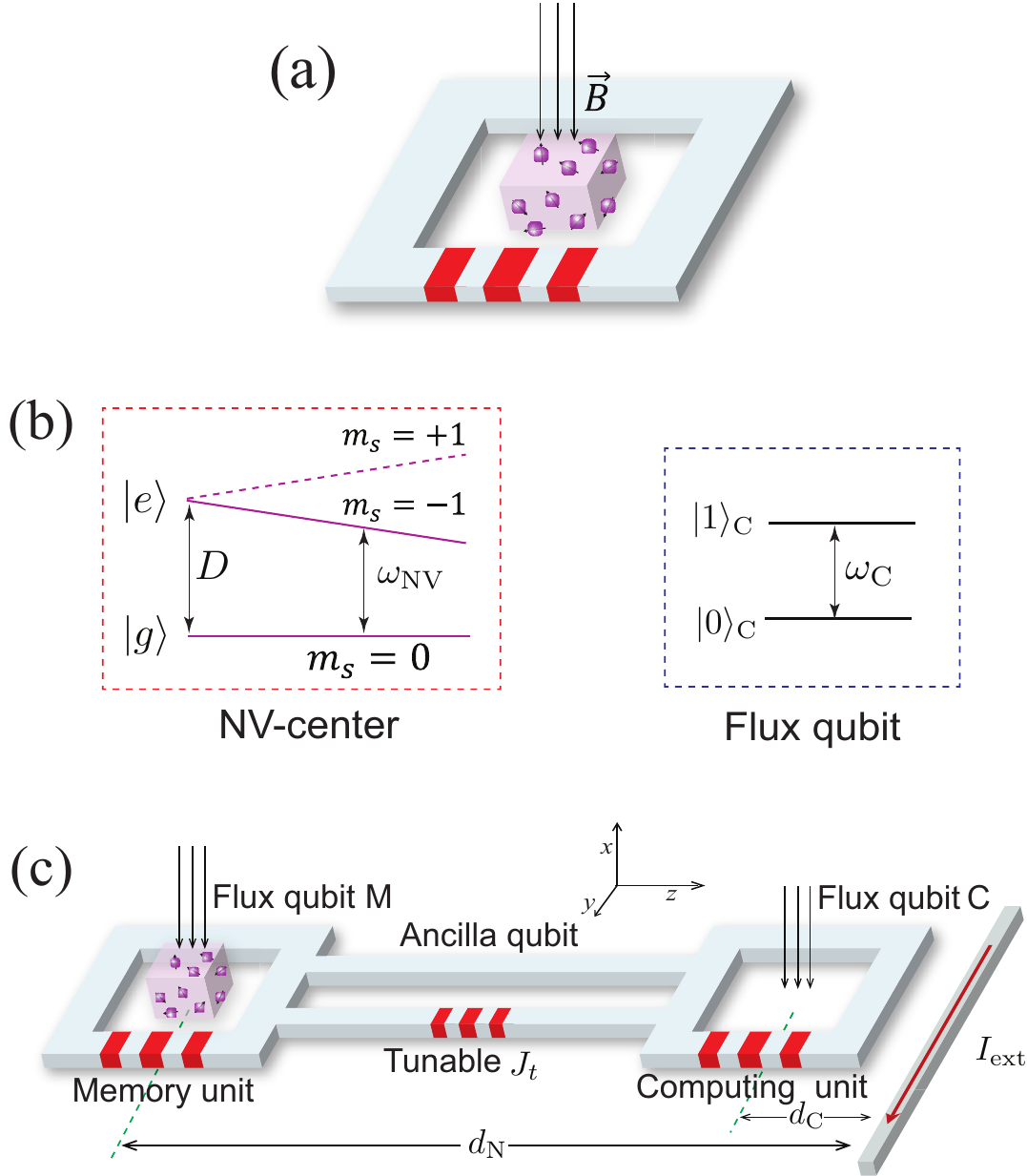}
\caption{ (Color online) (a) Single coupled flux-qubit-NVE system.  (b) Separate energy level structures of the NV center and flux qubit. (c)  Quantum memory circuit with separate memory and computing units. Two flux qubits are coupled with a tunable coupling strength $J_{t}$, and the right flux qubit serves as the computing unit. The NVE coupled to the left flux qubit serves as the spin-based quantum memory unit. Here $I_{\rm ext}$ denotes the external driving current separated from the flux qubit C and the NVE by the distances $d_{\rm C}$ and $d_{\rm N}$, respectively. By separating the computing and memory units, the influence of the quantum computing process on the quantum memory can be effectively eliminated, and hence the quantum storage of an arbitrary quantum state of the computing qubit could be achieved with high fidelity.}\label{fig1}
\end{figure}
As shown in Fig.\,\ref{fig1}, we consider a hybrid quantum circuit consisting of two coupled flux qubits (qubit M and C) and a NVE. The eigenstates of the flux qubit are superpositions of clockwise and counterclockwise persistent current states, and the energy levels are typically separated by a few GHz \cite{13}. 
An NV center has an $S=1$ ground state with zero-field splitting $D=2.88$ GHz between the $m_{s}=0$ and
$m_{s}=\pm1$ states. By introducing an external magnetic field along the crystalline axis of the NV center, an additional Zeeman splitting between $m_{s}=\pm1$ sublevels occurs. Thus, a two-level system with sublevels $m_{s}=0$ and $m_{s}=-1$ [see Fig.\,\ref{fig1}{\color{red} (b)}] can be isolated. 

We will first derive the Hamiltonian of the hybrid system consisting of a flux qubit and an NVE [see Fig.\,\ref{fig1}{\color{red} (a)}], and then obtain the total Hamiltonian of the proposed hybrid quantum circuit in Fig.\,\ref{fig1}{\color{red}(c)}.
\subsection{Flux qubit coupled to the NVE}
In the memory unit proposed here, the NV center can be described by the Hamiltonian \cite{13.5}
\begin{align}
\label{eq1}
H_{\rm N}&=DS^{2}_{z}+E(S^{2}_{x}-S^{2}_{y})+g_{e}\mu_{B} \vec{B}\cdot\vec{S},
\end{align}
where $D$ is the ground-state zero-field splitting, $\vec{S}$ are the usual Pauli spin-1 operators, $E$ is the ground-state strain-induced splitting coefficient, $g_{e}$ is the ground state $g$-factor, and $\mu_{B}$ is the Bohr magneton. In this paper, we set $\hbar=1$. Furthermore, we consider the case where the strain-induced fine-structure splitting is negligible compared to the Zeeman splitting, i.e., $|E (S^{2}_{x}-S^{2}_{y} )|\ll|g_{e}\mu_{B}\vec{B}\cdot\vec{S}|$. 
Thus, the second term in $H_{\rm N}$ can be neglected here.

The flux qubit M can create a superposition state of clockwise and counterclockwise persistent currents in the qubit loop with hundreds of nano-Amperes. Hence the magnetic field associated with these electric currents enables a magnetic-dipole coupling to the electron spins in the NV center. Specifically, we set the crystalline axis of the NV centers as the $z$-axis, and apply an external magnetic field $\vec{B}_{\rm ext}$. The component of $\vec{B}_{\rm ext}$ parallel to the $z$-axis tunes the NV centers into near-resonance with the flux qubit, and the component perpendicular of $\vec{B}_{\rm ext}$ to the qubit loop adjusts the superposition state of the clockwise and counterclockwise persistent-current states of the flux qubit. These two persistent-current quantum states of the flux qubits give rise to different anti-aligned magnetic fields: $\sigma_{\rm M}^{z}\vec{B}_{\rm FQ}$. Then we can express the Hamiltonian of the NVE coupled to the flux qubit M as
\begin{align}
\label{eq2}
H_{\rm NM}=&\frac{1}{2}(\varepsilon_{\rm M}\sigma^{z}_{\rm M}+\lambda_{\rm M}\sigma^{x}_{\rm M})+\sum^{N}_{k=1}\left[D(S^{k}_{z})^2\nonumber\right.
\\
&\left.+g_{e}\mu_{B}B^{\rm ext}_{z}S^{k}_{z}+\sigma_{\rm M}^{z}g_{e}\mu_{B}\vec{B}^{k}_{\rm FQ}\cdot\vec{S}^{k}\right],
\end{align}
where $\vec{\sigma}_{\rm M}$ denotes the Pauli operators of the flux qubit M; $\lambda_{\rm M}$ is the tunneling energy between the two wells of the qubit potential, and 
$\varepsilon_{\rm M}=2I_{p}(\Phi_{\rm M}-\Phi_{0}/2)$ is the energy bias of the flux qubit, with $I_{p}$ being its persistent current, $\Phi_{\rm M}$ the applied magnetic flux, and $\Phi_{0}$ the magnetic-flux quantum. Here $B^{\rm ext}_{z}$ is the part of the external magnetic field paralleled to the $z$-axis, which adjusts the energy splitting of the NV centers. Here, we assume that the $z$-axis is parallel to the surface of the flux qubits and make the two-level approximation for the NV centers. The Hamiltonian (\ref{eq2}) can then be rewritten as
\begin{align}
\label{eq3}
H_{\rm NM}=&\frac{1}{2}(\varepsilon_{\rm M}\sigma^{z}_{\rm M}+\lambda_{\rm M}\sigma^{x}_{\rm M})\nonumber
\\&+\omega_{\rm NV}\sum^{N}_{k=1}\tau_{k}^{+}\tau_{k}^{-}+\sum^{N}_{k=1}g_{k}\sigma^{z}_{\rm M}\left(\tau^{+}_{k}+h.c.\right).
\end{align}
Here $\tau$ denotes the Pauli operators of states with $m_{s}=0$ and $-1$ of the NV centers, and $\omega_{\rm NV}=D-g_{e}\mu_{B}B^{\rm ext}_{z}$ is the energy gap between these two states. The coupling strength $g_{k}$ between the flux qubit M and the individual spin is usually proportional to the magnitude of the qubit field at the spin location \cite{11,12}. The spins are assumed to have the homogeneous energy splitting $\omega_{\rm NV}$. We next introduce the collective operator $$b^{\dag}=g^{-1}\sum^{N}_{k}g_{k}\tau^{+}_{k},\;\;\;\;\;\;{\rm with}\;\;\;\;\;\;\; g=\left(\sum^{N}_{k}|g_{k}|^2\right)^{1/2},$$ and its Hermitian conjugate $b$. In the low-polarization limit, where almost all spins are in the ground
state, $b^{\dag}$ and $b$ obey approximately bosonic commutation relations, $[b, b^{\dag}]\approx1$ and
$\sum^{N}_{k}\tau^{+}_{k}\tau^{-}_{k}=b^{\dagger}b$ \cite{12.2}. Thus, the NVE can be considered to be an effective bosonic mode and the
interaction between the flux qubit M and the NVE becomes \cite{13.6,13.7,14},
\begin{align}
\label{eq4}
H^{\rm int}_{\rm NM}=g\sigma_{\rm M}^{z}\left(b^{\dag}+b\right).
\end{align}
Here, we only consider the two lowest state of mode $b$, $$|0\rangle_{\rm D}=|g_{1}g_{2}...g_{\rm N}\rangle$$ and $$|1\rangle_{\rm D}=\frac{1}{\sqrt{N}}\sum^{N}_{k}|g_{1}...e_{k}...g_{\rm N}\rangle.$$ The collective coupling $g$ is enhanced by a factor of $\sqrt{N}$ compared to the root-mean-square of the individual couplings $g_{k}$. Recent experiments achieved a coupling strength as strong as
$g\approx2\pi\times35$ MHz \cite{12.5}. 
\subsection{Total Hamiltonian of the hybrid quantum circuit}
In order to connect the computing and memory units in our proposal, the flux qubit M is also needed to couple the computing unit (flux qubit C) with a tunable coupling strength $J_{t}$. This tunable couping can be realized by a SQUID or ancilla flux qubit \cite{14.5,15,16,16.1,16.4,16.5}, and it also has become experimentally feasible to manipulatie this coupling strengths {\it in situ} and engineer various types of
circuit connectivities \cite{17,18,20}. Here, let us consider a general model for   
realizing this controllable coupling with an ancilla flux qubit, as shown in the Fig.\,\ref{fig1}{\color{red} (c)}.
By adiabatically eliminating the degrees of freedom of the ancilla flux qubit, the total Hamiltonian of the flux qubits M and C can be
written as \cite{14.5,15,16}
\begin{align}
\label{eq5}
H_{\rm MC}=&\frac{1}{2}\sum_{j={\rm M,C}}\left(\varepsilon_{j}\sigma^{z}_{j}+\lambda_{j}\sigma^{x}_{j}\right)+J_{t}\sigma_{\rm M}^{z}\sigma_{\rm C}^{z},
\end{align}
where $J_{t}$ is the tunable coupling strength, which can be adjusted by varying the flux piercing the superconducting loop of the ancilla qubit.
Here $\varepsilon_{j}$, $\lambda_{j}$ have the same meaning as before.

Then, considering the above spin-flux and flux-flux couplings,
the total Hamiltonian of the hybrid
system proposed here is given by
\begin{align}
\label{eq6}
H_{\rm tot}=&\frac{1}{2}\sum_{j={\rm M,C}}\left(\varepsilon_{j}\sigma^{z}_{j}+\lambda_{j}\sigma^{x}_{j}\right)+\omega_{\rm NV}b^{\dagger}b\nonumber
\\
&+g\sigma^{z}_{\rm M}\left(\tau^{+}_{k}+\tau_{k}\right)+J_{t}\sigma_{\rm M}^{z}\sigma_{\rm C}^{z}.
\end{align}
At the flux degenerate point, i.e., $\Phi_{j}=\Phi_{0}/2$ for the flux qubits M and C, and using rotating-wave approximation (RWA), the total Hamiltonian
can be reduced as
\begin{align}
\label{eq7}
H_{\rm tot}=&\sum_{j={\rm M,C}}\omega_{j}\tilde{\sigma}^{+}_{j}\tilde{\sigma}^{-}_{j}+\omega_{\rm NV}b^{\dagger}b\nonumber
\\
&+g\left(\tilde{\sigma}^{+}_{\rm M}b+b^{\dagger}\tilde{\sigma}^{-}_{\rm M}\right)+J_{t}\left(\tilde{\sigma}_{\rm M}^{+}\tilde{\sigma}_{\rm C}^{-}
+\tilde{\sigma}_{\rm C}^{+}\tilde{\sigma}_{\rm M}^{-}\right),
\end{align}
where $\omega_{j}=\lambda_{j}$ denotes the frequency of flux qubit $j$, and $\vec{\tilde{\sigma}}$ denotes the Pauli operators expressed in the eigenvector basis of the flux qubits.    

\begin{figure}[here]
\centering
\includegraphics[width=0.5\textwidth]{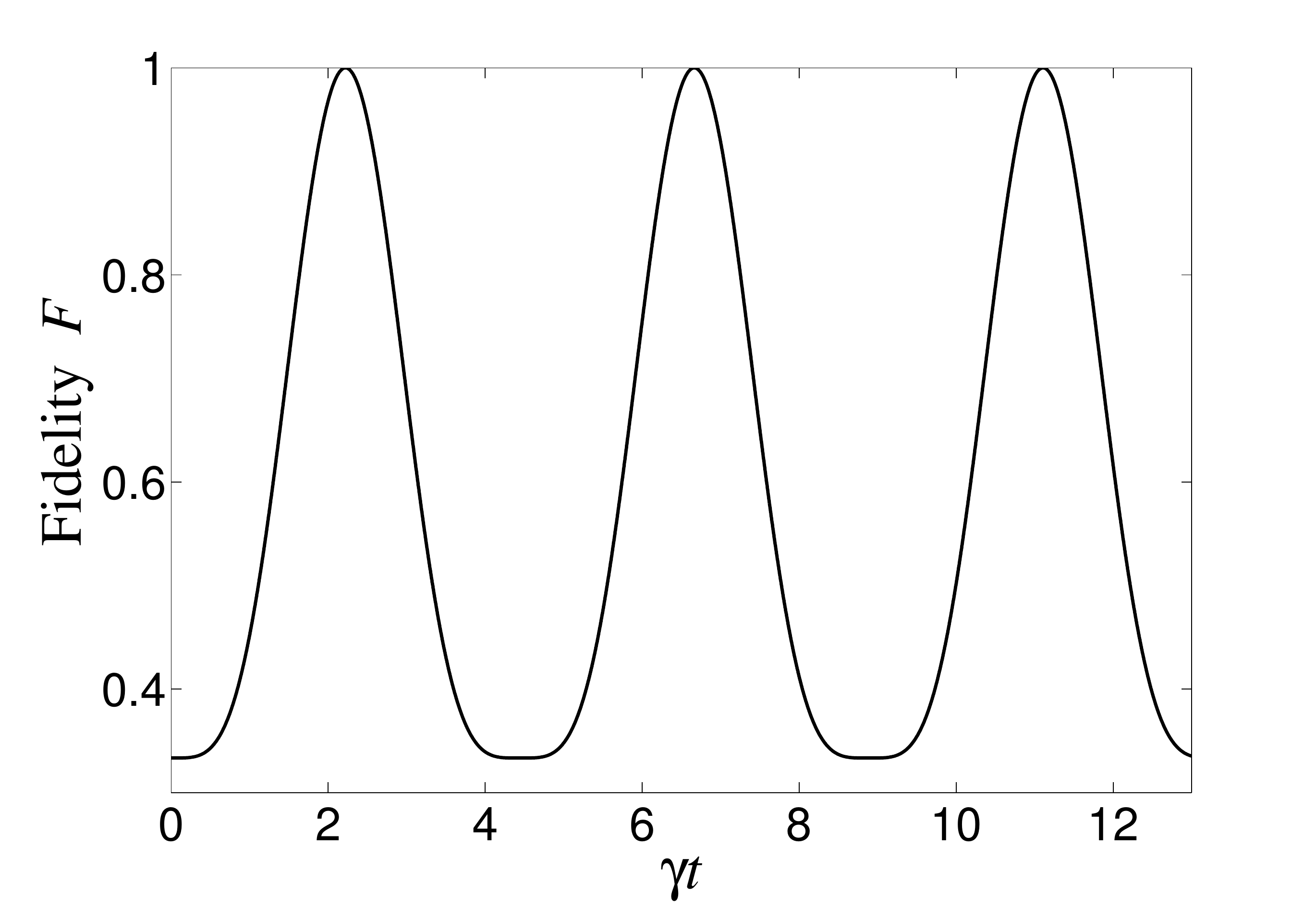}
\caption{The fidelity $F$ of quantum storage versus the dimensionless
time $\gamma t$ in the resonant-interaction case.
The parameters are scaled with $\gamma$ and chosen as
$\alpha=1/\sqrt{3}$, $\beta=\sqrt{2/3}$, $J_{t}=g=\gamma$, and
$\Delta_{\rm C}=\Delta_{\rm NV}=0$.}
\label{fig2}
\end{figure}

\begin{figure}[here]
\centering
\includegraphics[width=0.48\textwidth]{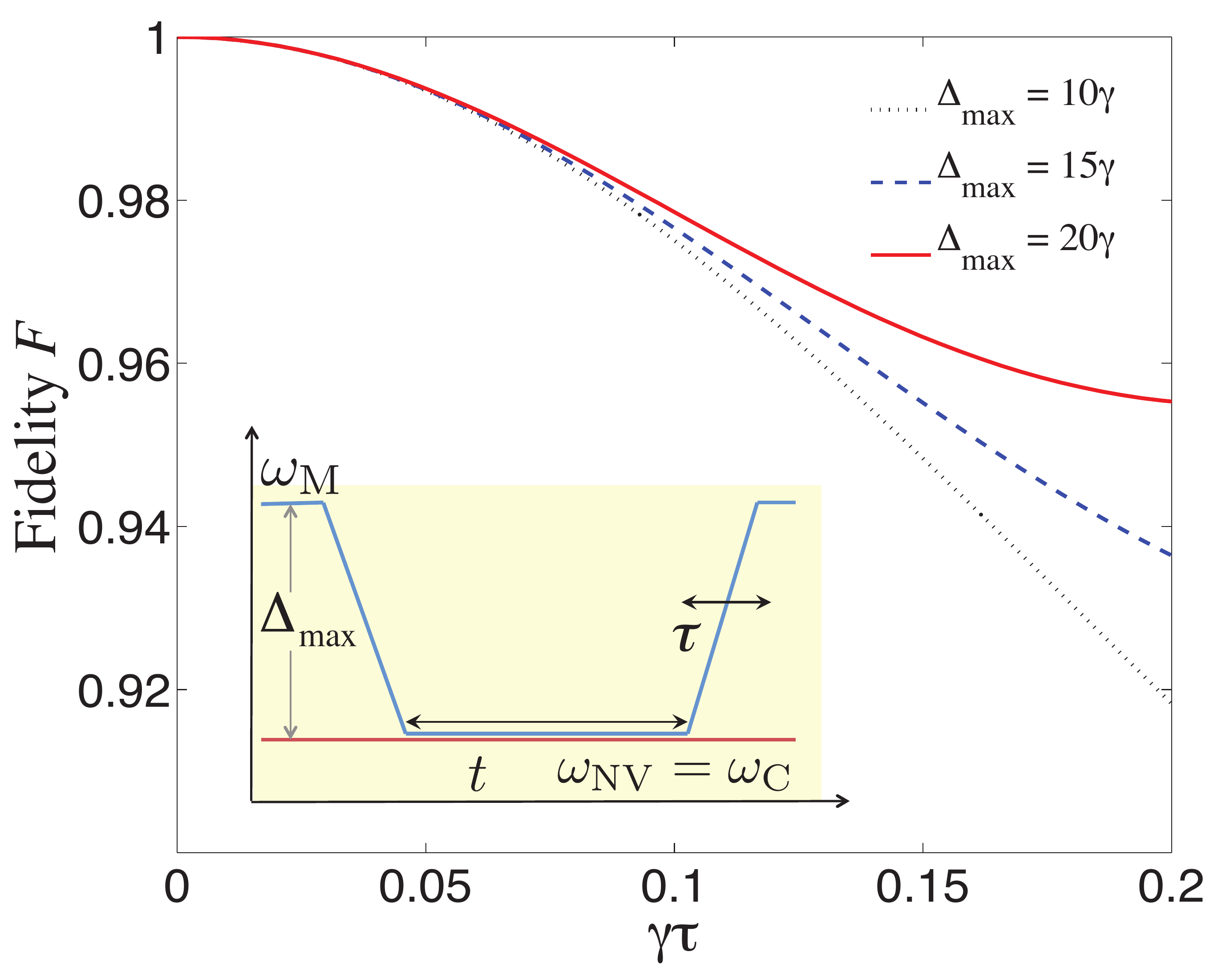}
\caption{(Color online) The maximum fidelity $F$ of quantum storage versus the dimensionless ramp time $\gamma\tau$ for different initial detuning $\Delta_{\rm max}=\omega_{\rm M}(0)-\omega_{\rm C}$ (or $\omega_{\rm M}(0)-\omega_{\rm NV}$ because here $\omega_{\rm C}=\omega_{\rm NV}$).The parameters are scaled with $\gamma$ and the same as in Fig.\,\ref{fig2}.}
\label{fig3}
\end{figure}

\begin{figure*}
\includegraphics[width=0.49\textwidth]{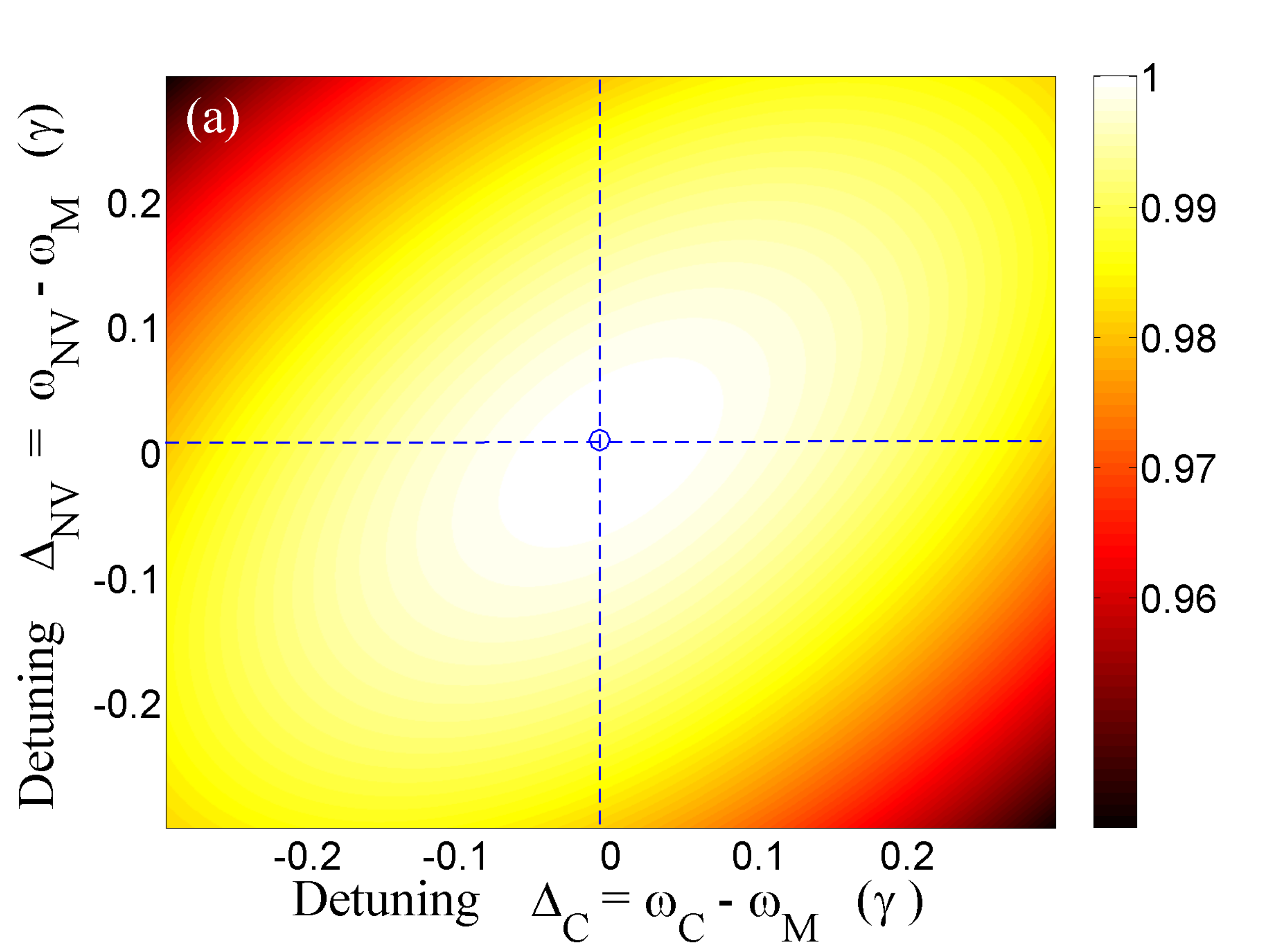}
\includegraphics[width=0.49\textwidth]{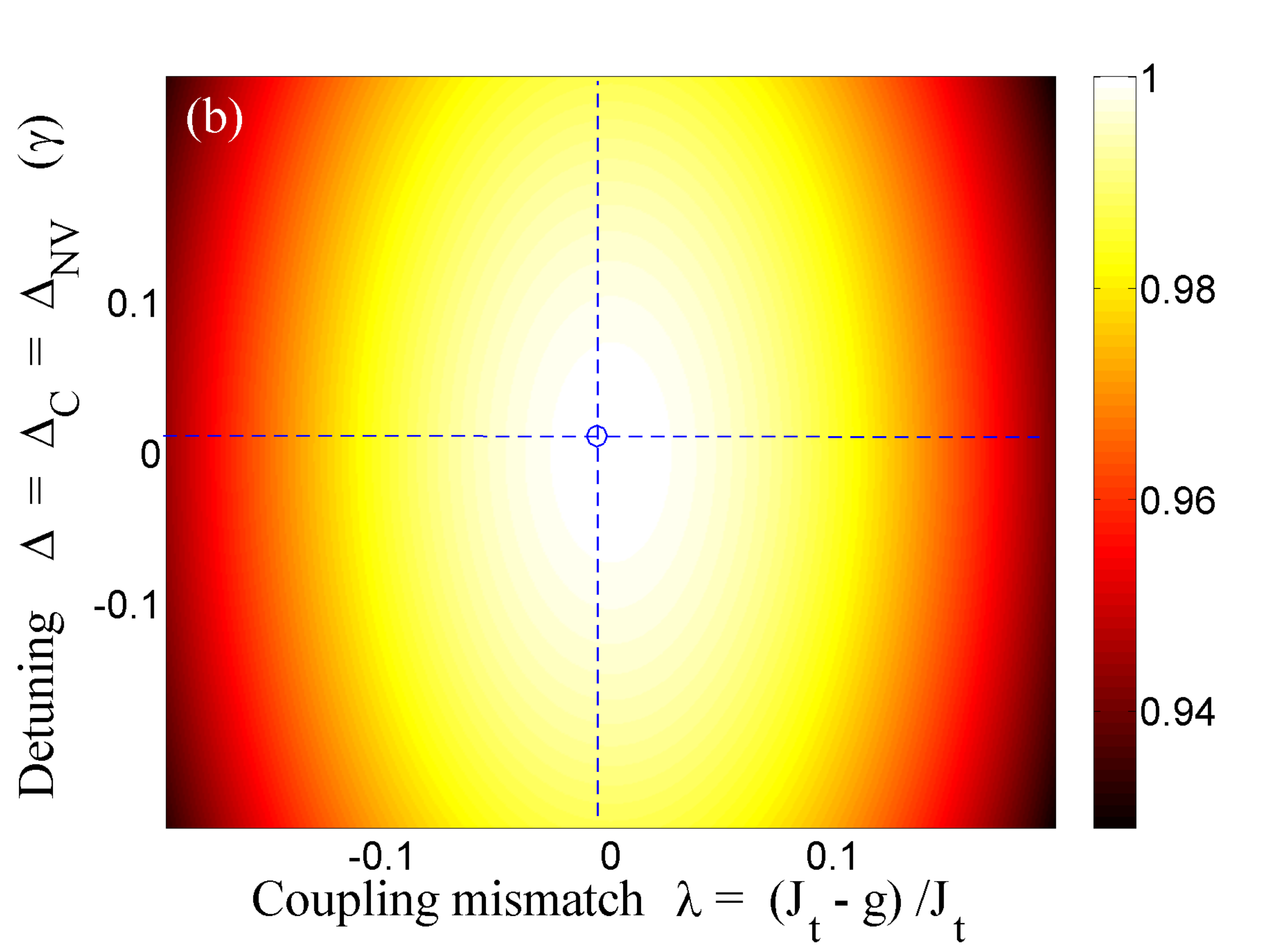}
\caption{(Color online) Fidelity $F$ of quantum storage versus: (a) the detunings $\Delta_{\rm C}$
and $\Delta_{\rm NV}$, (b) the detuning $\Delta$ (here $\Delta_{\rm C}=\Delta_{\rm NV}=\Delta$) and the dimensionless coupling mismatch $\lambda$.
The ideal parameter regime used in Fig.\,\ref{fig2} is indicated by the dashed
lines. The parameters are scaled with $\gamma$ and the same as in Fig.\,\ref{fig2}.}\label{fig4}
\end{figure*}
\section{Resonant interaction proposal for quantum storage}
In this section, we begin to discuss the quantum storage of an
arbitrary state of flux qubit C, that is
\begin{align}
|\psi_{I}\rangle=&\left(\alpha|0\rangle_{\rm C}+\beta|1\rangle_{\rm C}\right)|0\rangle_{\rm NV}\nonumber
\\
&\rightarrow|\psi_{T}\rangle=|0\rangle_{\rm C}\left(\alpha|0\rangle_{\rm NV}+\beta|1\rangle_{\rm NV}\right),\nonumber
\end{align}
based on the resonant interaction between the computing and memory units. Here, the frequencies of flux qubits ${\rm M}$ and ${\rm C}$ are  tunable, and we choose the qubit basis states, i.e., $|0\rangle_{\rm C}$, $|1\rangle_{\rm C}$, $|0\rangle_{\rm NV}$ and $|1\rangle_{\rm NV}$, as the computational basis.

In the resonant-interaction case, the frequencies
of two flux qubits and the NVE can be adjusted to satisfy the conditions $\Delta_{\rm NV}=\Delta_{\rm C}=0$, where the detunings are given by $$\Delta_{\rm NV}=\omega_{\rm M}-\omega_{\rm NV},\,\,\,\,\,\,\Delta_{\rm C}=\omega_{\rm M}-\omega_{\rm C}.$$ 
In the interaction
picture, we obtain the tripartite-resonant-interaction Hamiltonian \cite{12.6,21},
\begin{align}
\label{eq8}
H^{R}_{\rm tot}=g\left(b^{\dag}\tilde{\sigma}^{-}_{\rm M}+b\tilde{\sigma}^{+}_{\rm M}\right)+J_{t}\left(\tilde{\sigma}^{-}_{\rm C}\tilde{\sigma}^{\dag}_{\rm M}+\tilde{\sigma}^{+}_{\rm C}\tilde{\sigma}^{-}_{\rm M}\right).
\end{align}
We consider the computing qubit C to be initially in an
arbitrary state $\alpha|0\rangle_{\rm C}+\beta|1\rangle_{\rm C}$, and both the
flux qubit M and the NVE in the ground state
$|0\rangle_{\rm M}|0\rangle_{\rm NV}$. Then, the initial state of the system
$|\psi(0)\rangle$ is the coherent superposition state
$(\alpha|0\rangle_{\rm C}+\beta|1\rangle_{\rm C})|0\rangle_{\rm M}|0\rangle_{\rm NV}$.
The system state evolves following
\begin{align}
\label{eq9}
i\frac{\partial}{\partial
t}|\psi(t)\rangle=H^{R}_{\rm tot}|\psi(t)\rangle.
\end{align}
In the subspace
$\{|\phi_{1}\rangle=|1\rangle_{\rm C}|0\rangle_{\rm M}|0\rangle_{\rm NV};
|\phi_{2}\rangle=|0\rangle_{\rm C}|1\rangle_{\rm M}|0\rangle_{\rm NV};
|\phi_{3}\rangle=|0\rangle_{\rm C}|0\rangle_{\rm M}|1\rangle_{\rm NV}\}$. When $|\psi(0)\rangle=|1\rangle_{P}|0\rangle_{\rm M}|0\rangle_{\rm NV}$ we can
obtain the state of the system at time $t$,
$|\psi(t)\rangle=\sum^{3}_{j=1}C_{j}|\phi_{j}\rangle$ with
\begin{subequations}
\begin{align}
\label{eq10}
C_{1}&=J_{t}^{2}{\rm
cos}(\sqrt{J_{t}^{2}+g^{2}}t)/\left(J_{t}^{2}+g^{2}\right),
\\
C_{2}&=-iJ_{t}{\rm sin}(\sqrt{J^{2}+g^{2}}t)/\sqrt{J^{2}+g^{2}},
\\
C_{3}&=Jg\left[{\rm
cos}(\sqrt{J^{2}+g^{2}}t)-1\right]/\left(J^{2}+g^{2}\right).
\end{align}
\end{subequations}

From the above equations, we notice that $|\psi(t)\rangle=|0\rangle_{\rm C}|0\rangle_{\rm M}|1\rangle_{\rm NV}$ when
$\sqrt{2}gt=(2k+1)\pi$, $(k=0,1,2...)$, with parameter condition
$J_{t}=g$. When $|\psi(0)\rangle=|0\rangle_{\rm C}|0\rangle_{\rm M}|0\rangle_{\rm NV}$, the
system state will remain unchanged with time. Thus, the quantum storage process,
$(\alpha|0\rangle_{\rm C}+\beta|1\rangle_{\rm C})|0\rangle_{\rm NV}\rightarrow|0\rangle_{\rm C}(\alpha|0\rangle_{\rm NV}+\beta|1\rangle_{\rm NV})$
can be realized perfectly in the resonant interaction case. In order to further explicitly show the generation of high-fidelity
quantum storage, we plot in Fig.\,\ref{fig2} the quantum-storage-fidelity versus the dimensionless time $\gamma t$. Here the fidelity is defined as $F=|_{\rm M}\langle0|\langle\psi_{T}|\psi(t)\rangle|^{2}$ (here $|\psi_{T}\rangle$ is the target state of quantum storage) and this figure
shows that the quantum storage process can be deterministically realized at an appropriate time. In other words, the maximum fidelity for implementing quantum storage can reach up to one in the parameter regime: $\Delta_{\rm C}=\Delta_{\rm NV}=0$, and $g=J_{t}$.

Notice that the above discussion is in the ideal situation, where the corresponding undesired transitions from the computation basis to other subspace are neglected and the frequencies of the flux qubits are fixed. Experimentally, the above quantum storage is implemented by tuning the frequencies of the flux qubits. First, the flux qubit ${\rm C}$ is initially resonant with the NVE, but largely detunes from the flux qubit ${\rm M}$, i.e., $\omega_{\rm M}(0)\gg\omega_{\rm C},\omega_{\rm NV}$. In this situation, the interactions between the flux qubits ${\rm C}$, ${\rm M}$ and the NVE are negligible when flux qubit ${\rm C}$ and the NVE are sufficiently apart from each other. Second, we prepare the flux qubits ${\rm C}$, ${\rm M}$ and the NVE to their initial states, and bring the flux qubit ${\rm M}$ in resonance with flux qubit ${\rm C}$ and the NVE during a ramp time $\tau$ using a pulse, i.e., $\omega_{\rm M}(\tau)=\omega_{\rm C}=\omega_{\rm NV}$ (the ideal situation corresponding to $\tau\rightarrow0$). See the inset of Fig.\,3. Finally, we wait for a specific time as shown above, and then bring the qubit ${\rm M}$ back to the initial detuned position with another pulse \cite{21.5}.

Summing up the above discussion, the ideal parameter regime, i.e., $\tau\rightarrow0$, $\Delta_{\rm C}=\Delta_{\rm NV}=0$, and $g=J_{t}$, is necessary in order to obtain the perfect quantum storage process. However, the idea parameter conditions could not be
satisfied exactly in practical situations. In order to study the
influences of parameter mismatches on the fidelity of quantum storage, we plot $F$ against the ramp time of pulse $\tau$, detunings $\Delta_{\rm C}$, $\Delta_{\rm NV}$
and the coupling mismatch $$\lambda=(J_{t}-g)/J_{t}$$ in Figs.\,\ref{fig3} and \ref{fig4}. Figure\,\ref{fig3} shows that the quantum-storage-fidelity is robust with the ramp time $\tau$, and high-fidelity quantum storage can still be obtained even for a finite ramp time. For example, by choosing the flux-NVE coupling strength $g/2\pi=35$ MHz, the fidelity of quantum storage can still reach up to $97.5\%$ when the ramp time $\tau\approx0.45$ ns and the initial frequency detuning $\Delta_{\rm max}/2\pi=700$ MHz. Furthermore, it is clearly shown from Fig.\,\ref{fig4} that the quantum-storage-fidelity is
insensitive with respect to the fluctuations of parameters $\Delta_{\rm C}$, $\Delta_{\rm NV}$, and $\lambda$. As a result, the quantum
storage process can still be realized with high fidelity even though the ideal parameter conditions could not be satisfied exactly in
practical situations.

Here it should be noticed that the quantum storage process can also be realized in a single flux-qubit-NVE system \cite{11}. However, in the single flux-qubit-NVE system, it is necessary to apply additional pulse sequences on the memory (or computing) unit in order to eliminate the influence of the quantum computing process on the quantum memory. This requirement increases the difficulty of experimentally realizing a high-fidelity quantum storage. Next, we will show that the above problem can be solved by separating the quantum computing and memory units in our proposal (see Fig.\,\ref{fig1}{\color{red}(c)}). 

As shown in Fig.\,\ref{fig1}{\color{red}(c)}, an external current $I_{\rm ext}$ is used to perform the quantum computing process. The magnetic field strength at a distance $d$ away from the external current $I_{\rm ext}$ is
\begin{align}
\label{eq11}
\vec{B}(d)&=\mu_{0}\vec{I}_{\rm ext}/2\pi d,
\end{align}
where $\mu_{0}$ is the permeability of the vacuum. To estimate the coupling strength $\Omega_{\rm C}$ between the external magnetic field and the flux qubit C, we note that $H_{\rm BC}=-\vec{\mu}\cdot\vec{B}$, where $\vec{\mu}$ is the magnetic dipole of the flux qubit induced by the circulating persistent current of magnitude $I_{\rm C}$ and $\mu=I_{\rm C}A_{\rm C}$, where $A_{\rm C}=L^{2}$ is the area of the flux qubit C. When the frequency of the external magnetic $\omega_{d}$ is resonant with the transition frequency $\omega_{\rm C}$, this coupling Hamiltonian can be written as $H_{\rm BC}=\Omega_{\rm C}(\tilde{\sigma}^{+}_{\rm C}+\tilde{\sigma}^{-}_{\rm C})$, where 
\begin{align}
\label{eq12}
\Omega_{\rm C}&=\frac{\mu_{0}A_{\rm C}I_{\rm C}I_{e}}{2\pi\hbar d_{\rm C}}
\end{align}
is the Rabi frequency.

Similarly, the interaction of the external magnetic field $\vec{B}$ with the NV center can be written as $\vec{S}\cdot\vec{W}$, where $\vec{W}=g_{e}\mu_{B}\vec{B}$, $g_{e}$ and $\mu_{B}$ have the same meaning as in section II. Considering the magnetic field $\vec{B}$ along the $x$ axis (perpendicular to the crystalline axis of the NV center) and applying the RWA, the interaction Hamiltonian between the external magnetic field and the NVE can be written as 
$H_{\rm BN}=\Omega_{\rm NV}(b+b^{\dagger})$, where 
\begin{align}
\label{eq13}
\Omega_{\rm NV}=\sqrt{N}\frac{g_{e}\mu_{B}\mu_{0}I_{e}}{2\hbar\pi d_{\rm N}}.
\end{align}
Here the resonant condition between the flux qubit C and the NV-center has been used.
Based on the above discussion, it should be noticed that the external magnetic field will influence the quantum memory unit (NVE) when a single qubit rotation is applied on the quantum computing unit (flux qubit C). As an example, we consider that a single qubit operation (rotating $\theta$ angle) on the flux qubit C is firstly implemented by the external current $I_{\rm ext}$, and then the generated quantum state of the flux qubit C is transferred into the NVE. In particular, a small quantum state rotation for the NVE, $$|0\rangle_{\rm NV}\,\rightarrow\,\cos(\Omega_{\rm NV}t)|0\rangle-i\sin(\Omega_{\rm NV}t)|1\rangle_{\rm NV}$$ will occur when single quantum rotation $$|0\rangle_{\rm C}\,\rightarrow\,\cos(\Omega_{\rm C}t)|0\rangle_{\rm C}-i\sin(\Omega_{\rm C}t)|1\rangle_{\rm C}\;\;\;\;\;\;(\theta=\Omega_{\rm C}t)$$ is applied on the flux qubit C. As shown in Eq.\,(\ref{eq13}), the Rabi frequency $\Omega_{\rm NV}$ decreases, when increaseing the distance $d_{\rm N}$. Therefore, the influence of single-qubit-rotations on the memory unit could be effectively suppressed by separating the quantum memory and computing units because $\sin(\Omega_{\rm NV}t)\rightarrow0$ when $\Omega_{\rm NV}\rightarrow0$. A more detailed study on this will be presented in section V by numerically simulating the quantum-storage-fidelity using typical actual experimental parameters.  

Before ending this subsection, we would like to point out another advantage of our proposal. Here, the coupling between two flux qubits can be easily controlled by an external dc magnetic field. Thus, we can cutoff (or turn on) the connection between the quantum computing and memory units by a dc magnetic field when quantum gate operations (or quantum information transfer) are implemented. This property ensures that the reversible quantum storage process can be easily realized in our proposal without needing any additional operations applied on either the computing or memory units. 
 
\section{Dispersive interaction proposal for quantum memory}
\begin{figure}[here]
\centering
\includegraphics[width=0.51\textwidth]{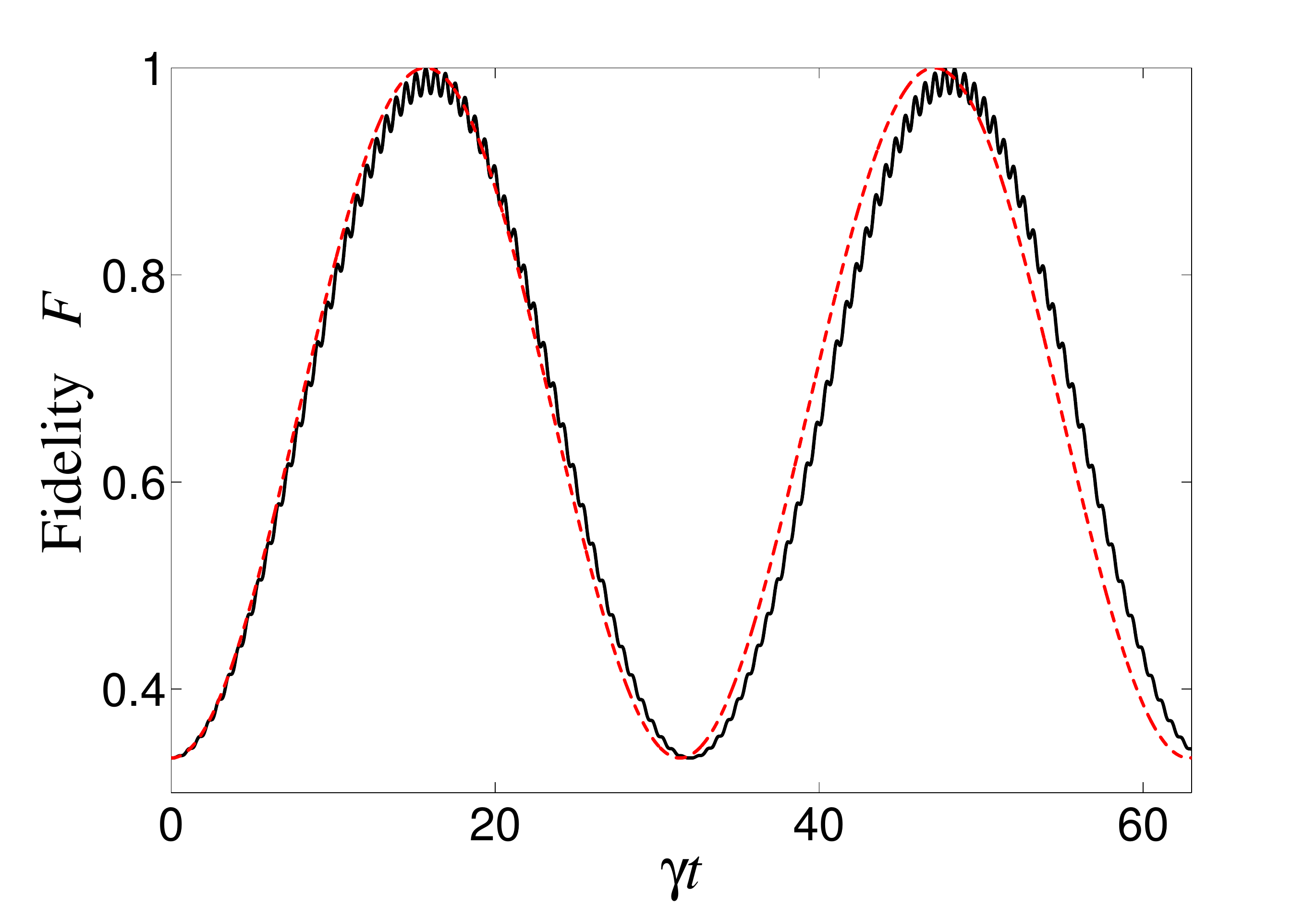}
\caption{(Color online) The fidelity $F$ of quantum storage versus the dimensionless time $\gamma t$ in the dispersive interaction case. The black solid and red dashed cuvers correspond to the original Hamiltonian (\ref{eq7}) and the effective Hamiltonian (\ref{eq14}), respectively. The parameters are scaled with $\gamma$ and the same as in Fig.\,\ref{fig2}, except for $\Delta_{\rm C}=\Delta_{\rm NV}=10\gamma$.}\label{fig5}
\end{figure}
\begin{figure}
\includegraphics[width=0.48\textwidth]{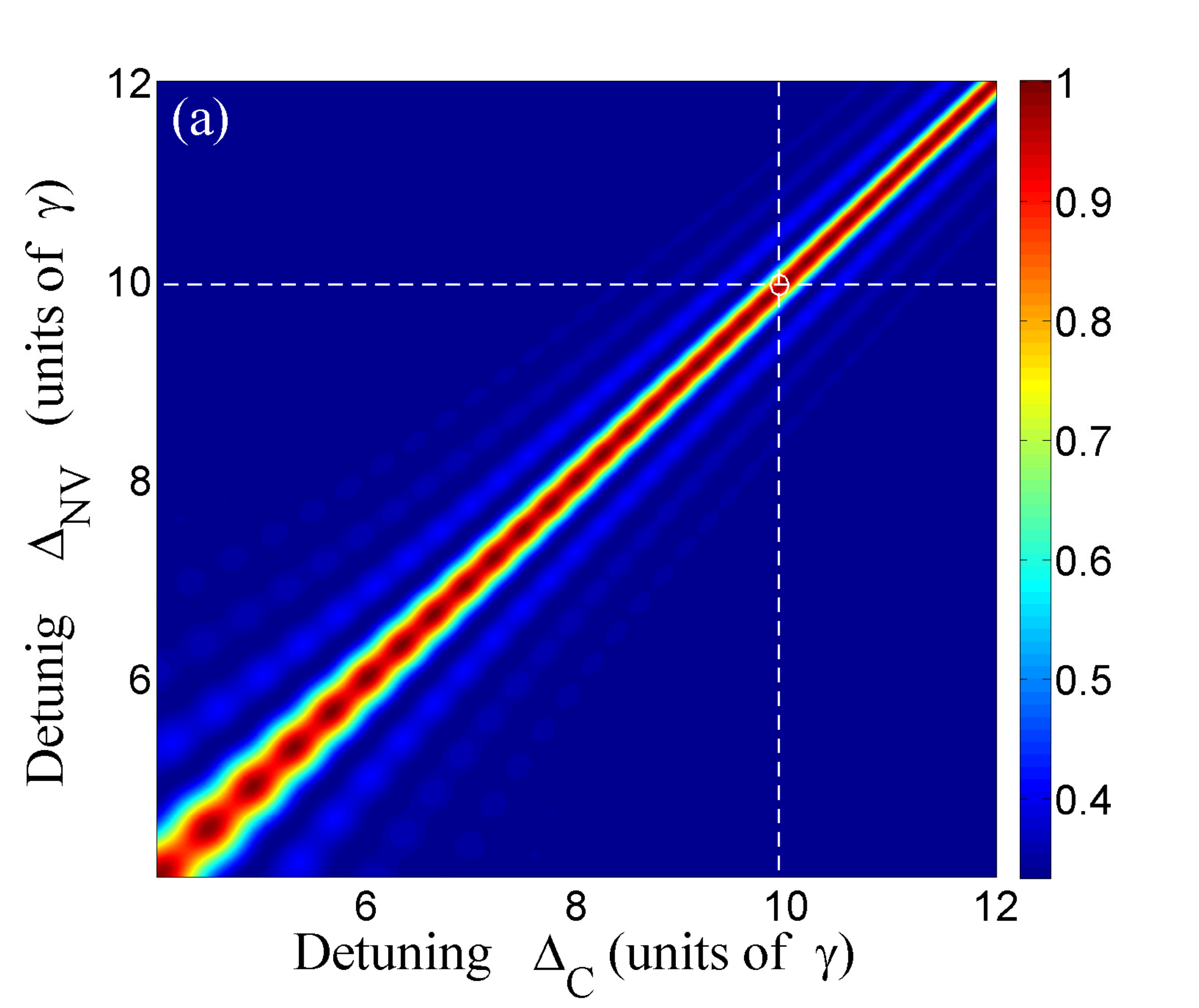}
\includegraphics[width=0.48\textwidth]{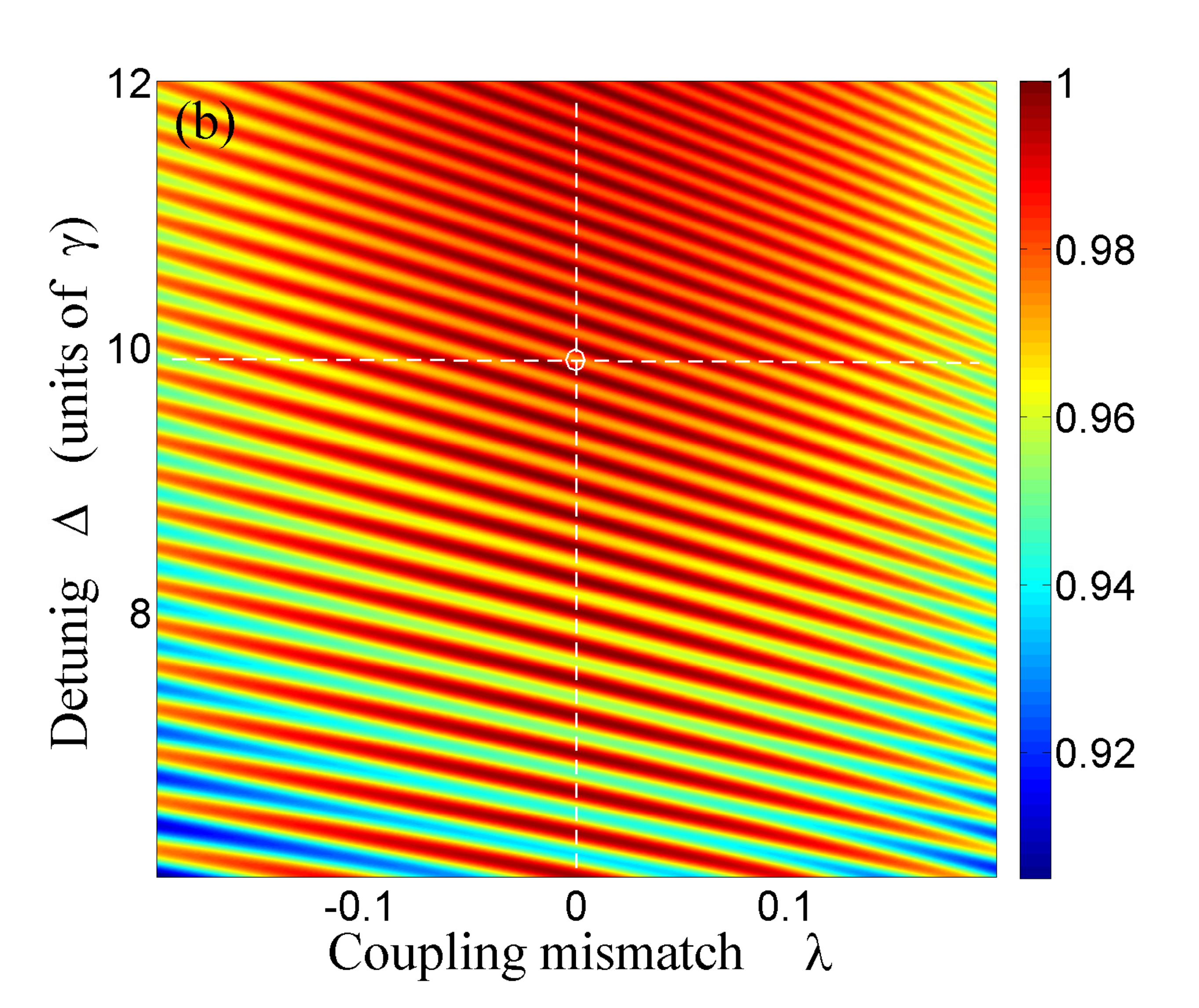}
\caption{(Color online) The fidelity $F$ of the quantum storage versus: (a) the detunings $\Delta_{\rm C}$
and $\Delta_{\rm NV}$, and (b) the detuning $\Delta$ (here $\Delta_{\rm C}=\Delta_{\rm NV}=\Delta$) and the dimensionless mismatch $\lambda$. The ideal
parameter condition used in Fig.\,\ref{fig5} are indicated by the dashed
lines. The parameters are scaled with $\gamma$ and the same as in Fig.\,\ref{fig5}.}\label{fig6}
\end{figure}

In this subsection, we calculate the evolution of the system in the dispersive interaction case and show the realization of high-fidelity quantum storage. In the dispersive-interaction case, the
frequencies of the flux qubit C and the NVE are
detuned from the frequency of qubit M by $\Delta_{\rm NV}$, and
$\Delta_{\rm C}$ (where $\Delta_{\rm NV}, \Delta_{\rm C}\gg g,
J_{t}$). Then, we can adiabatically eliminate the degree of freedom of the qubit M and obtain the Hamiltonian of the effective interaction between the NVE and flux qubit C by a Fr\"{o}hlich transformation \cite{22,23,24} 
\begin{align}
\label{eq14}
H^{D}_{\rm tot}&={\rm exp}(-S)H_{\rm tot}{\rm exp}(S)\nonumber
\\
&=\left(\Delta_{\rm NV}+\frac{g^{2}}{\Delta_{\rm NV}}\right)b^{\dag}b+\left(\Delta_{\rm C}+\frac{J_{t}^{2}}{\Delta_{\rm C}}\right)\tilde{\sigma}^{\dag}_{\rm C}\tilde{\sigma}^{-}_{\rm C}\nonumber
\\
  &\;\;\;\;\;+\Lambda\left(\tilde{\sigma}^{-}_{\rm C}b^{\dag}+\tilde{\sigma}^{\dag}_{\rm C}b\right),
\end{align}
where $$S=\frac{g}{\Delta_{\rm NV}}\left(\tilde{\sigma}^{-}_{\rm M}b^{\dagger}-b\tilde{\sigma}^{+}_{\rm M}\right)+\frac{J_{t}}{\Delta_{\rm C}}\left(\tilde{\sigma}^{-}_{\rm M}\tilde{\sigma}^{+}_{\rm C}-\tilde{\sigma}_{\rm C}^{-}\tilde{\sigma}^{+}_{\rm M}\right).$$
Here $$\Lambda=\frac{gJ_{t}}{2}\left(\frac{1}{\Delta_{\rm NV}}+\frac{1}{\Delta_{\rm C}}\right)$$ is the effective
coupling between the flux qubit C (computing unit) and the NVE (memory unit). In the above calculation, we have
assumed that the qubit M is initially prepared in its ground state.

From the Hamiltonian (\ref{eq14}), we show that the flux qubit C and
NVE will exchange energy by virtually exciting the qubit M, which
could effectively avoid the losses induced by qubit M. So, in the
dispersive interaction case, the goal of quantum storage, that is,
$$(\alpha|0\rangle_{\rm C}+\beta|1\rangle_{\rm C})|0\rangle_{\rm NV}\rightarrow|0\rangle_{\rm C}(\alpha|0\rangle_{\rm NV}+\beta|1\rangle_{\rm NV})$$
can also be realized perfectly at the appropriate time $\Lambda
t=(2k+1)\pi/2$, for the ideal parameter conditions:
$\Delta_{\rm C}=\Delta_{\rm NV}$ and $g=J_{t}$. 

In order to show the validity of the above discussion, we plot the fidelity $F$ of the quantum storage with the original Hamiltonian (\ref{eq7}) and the effective Hamiltonian (\ref{eq14}) in Fig.\,\ref{fig5}.  The consistency between the dashed and solid lines in this Figure proves that the high-fidelity quantum storage process can still be realized in the dispersive-interaction case.

In Fig.\,\ref{fig6}, we also present the influences of the system parameters on the quantum-storage-fidelity in the dispersive-interaction case.
Figure\,\ref{fig6}{\color{red}(a)} shows that the frequencies of the computing and memory qubits still need to be close, i.e., $\Delta_{\rm C}\simeq\Delta_{\rm NV}$, in order to obtain high fidelity. It can also be seen from Fig.\,\ref{fig6}{\color{red}(b)} that the fidelity of the quantum storage is robust with respect to the system parameter $\lambda$ for large detuning. Notice that the interference fringe in Fig.\,\ref{fig6} comes from the detuning-induced frequency-shift on the computing and memory qubits when we adiabatically eliminate the degrees of the flux qubit M. 

\section{Experimental Feasibility}
\begin{figure*}
\includegraphics[width=0.45\textwidth]{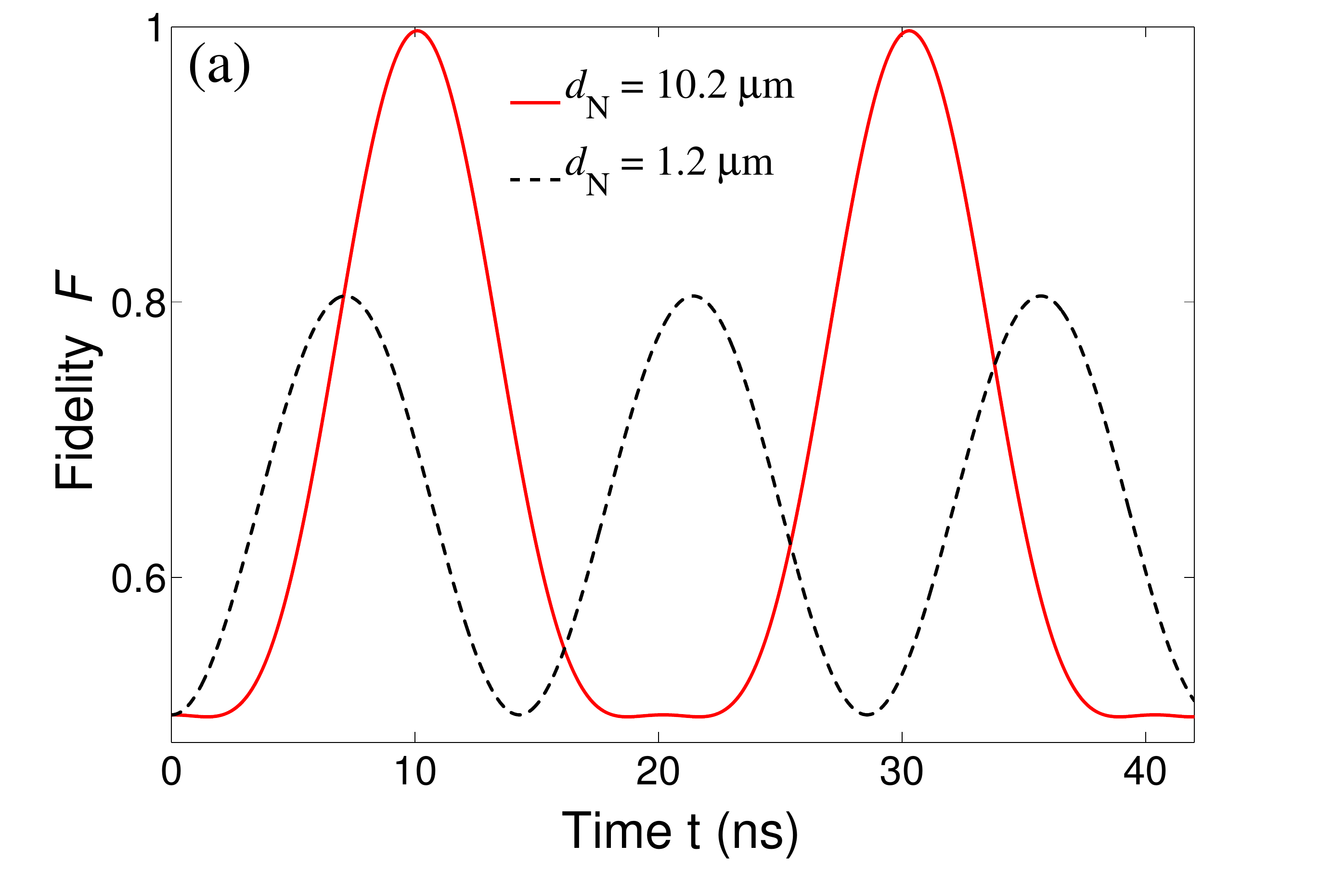}
\includegraphics[width=0.45\textwidth]{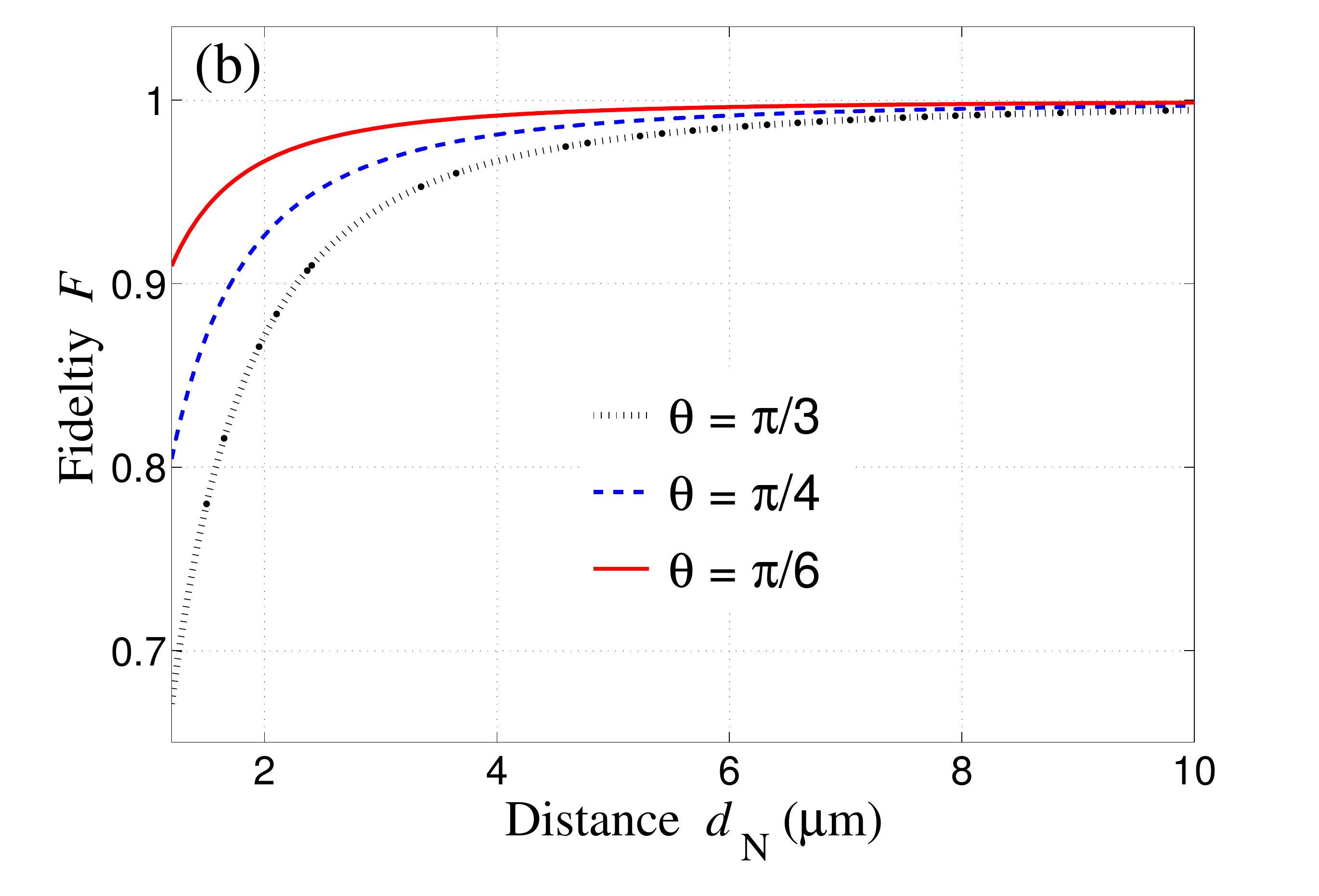}
\includegraphics[width=0.45\textwidth]{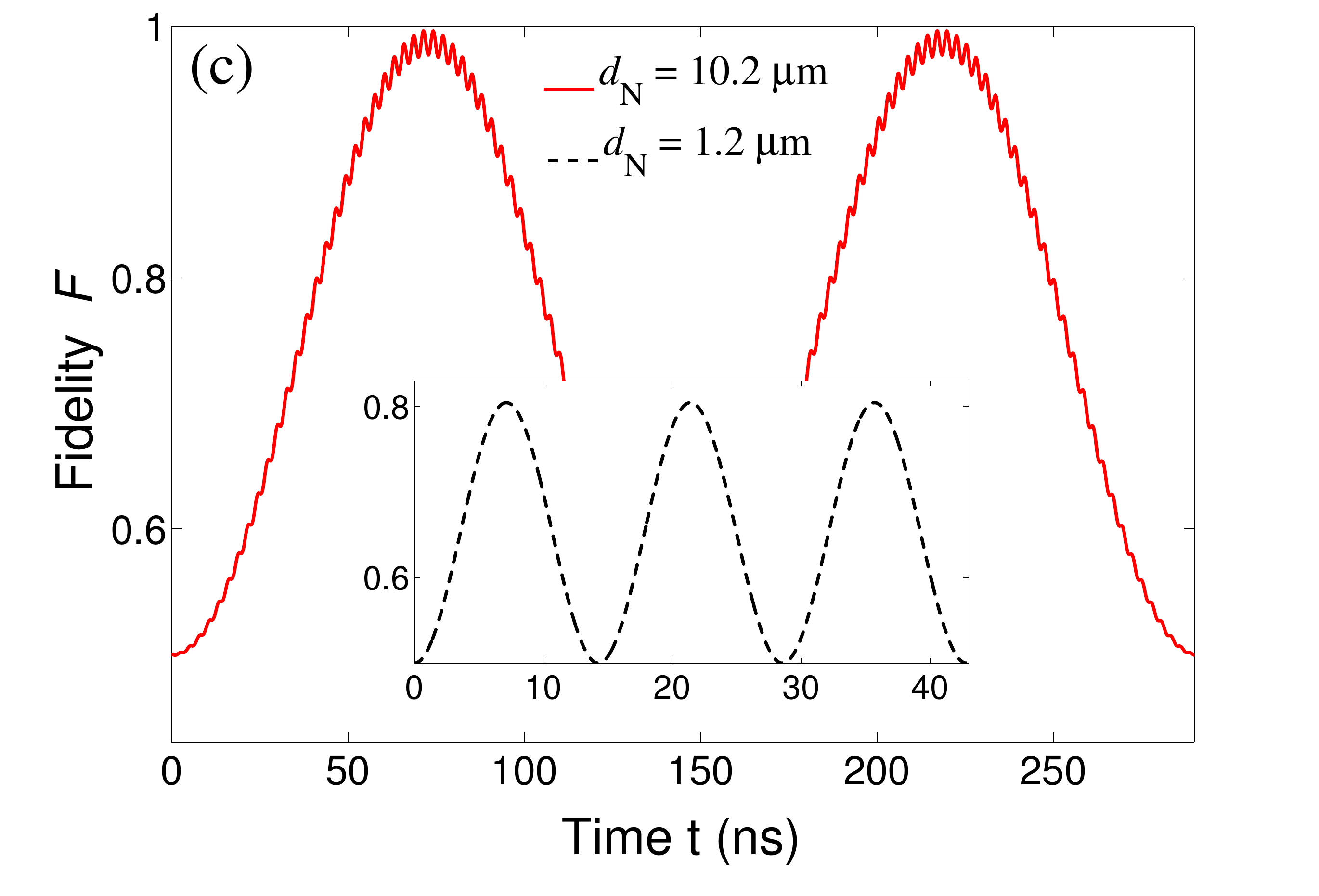}
\includegraphics[width=0.45\textwidth]{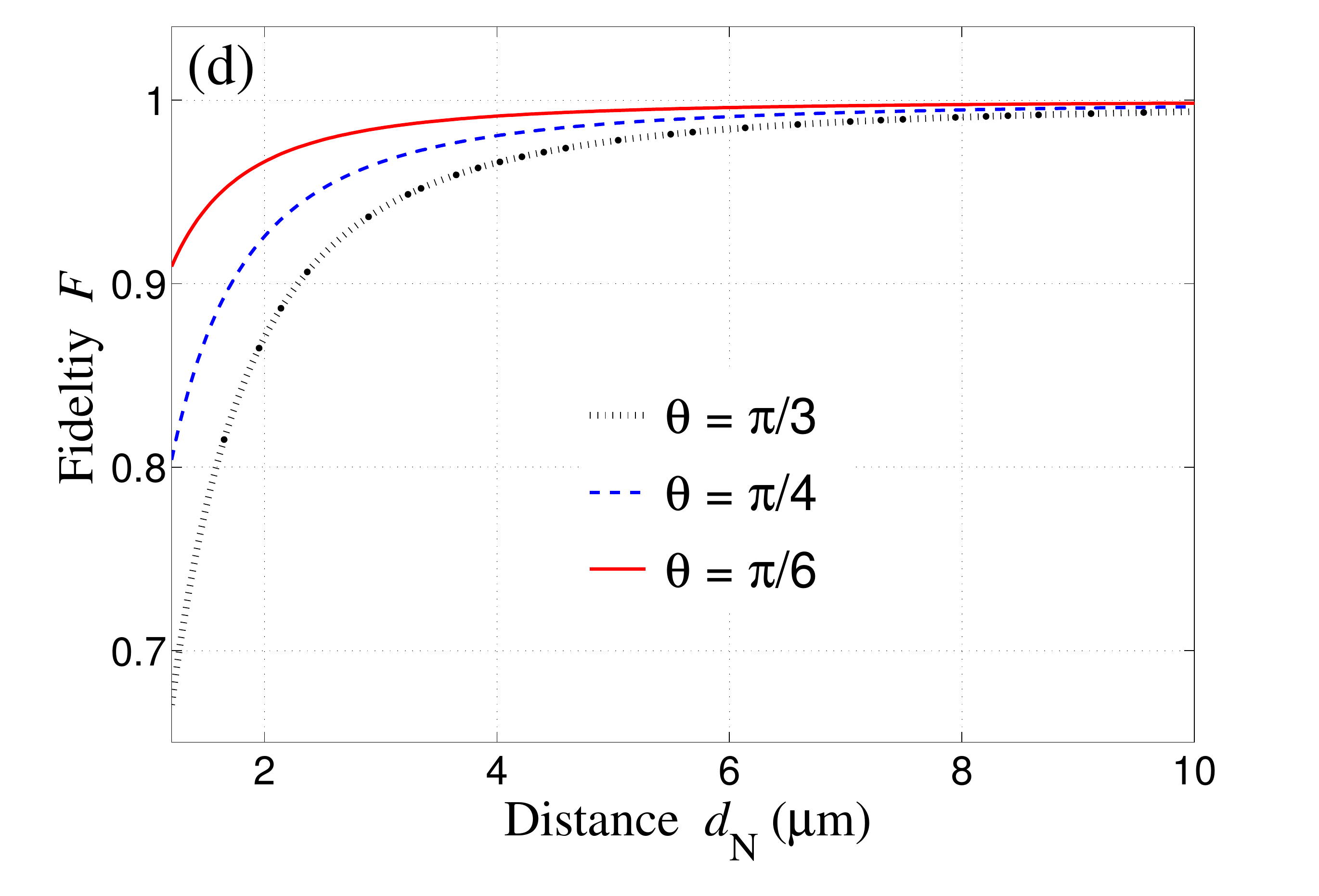}
\caption{(Color online) The fidelities of quantum storage versus: (a,c) time $t$ and (b,d) the distance $d_{\rm N}$ (shown in Fig.\,\ref{fig1}). The black dashed and red solid curves in (a,c) correspond to the single flux-qubit-NVE system and the proposed system in this paper. The system parameters used here are: $L_{\rm C}=2$ ${\rm \mu}$m, $N=10^{6}$, $d_{\rm C}=1.2$ ${\rm \mu}$m, $I_{\rm C}=60$ nA, $I_{e}=700$ nA, and $J_{t}/2\pi=g/2\pi=\Omega_{\rm C}/2\pi\approx$ 35 MHz. Figures (a,b) and (c,d) correspond to the case of resonant ($\Delta_{\rm C}=\Delta_{\rm NV}=0$) and dispersive ($\Delta_{\rm C}/2\pi=\Delta_{\rm NV}/2\pi=350$ MHz) interactions, respectively.  Notice that $\theta=\pi/4$ was chosen in (a,c) and the maximum fidelity $F$ was chosen in (b,d).}\label{fig7}
\end{figure*}

Let us now discuss the experimental feasibility of our proposal. To be consistent with the discussion in section III, in Fig.\,\ref{fig7} we calculate the quantum-storage-fidelities with actual experimental parameters based on the single flux-qubit-NVE system and the proposed system here, respectively. The dynamics of the single (hence the sub-index ``s" in $H_{s}$) flux-qubit-NVE system is decided by the interaction Hamiltonian $H_{s}=g(b^{\dagger}\tilde{\sigma}^{-}_{\rm C}+b\tilde{\sigma}^{+}_{\rm C})$. Figures 7(a,b) and (c,d) correspond to the case of resonant and dispersive interactions, respectively. It is clearly shown from Figs.\,\ref{fig7}{\color{red}(a) and (c)} that the influence of the quantum computing process on the quantum memory can be effectively suppressed by separating the computing and memory units. In addition, we also clearly show in Figs.\,\ref{fig7}{\color{red}(b) and (d)} the dependence of the quantum-storage-fidelity on the distance $d_{\rm N}$, and thus, a high-fidelity quantum storage process can be achieved in our proposal by choosing a proper distance ($d_{\rm N}>8$\,$\mu$m) between two flux qubits.  The present numerical results are consistent with the qualitative conclusion obtained in section III, and clearly show the feasibility of our proposal when using experimental parameters.

We list the following experimental conditions required by our proposal: 
(i) the flux qubits M and C should be connected with a tunable coupling strength, and have a long coherence time of the order of a microsecond;  
(ii) the density of the NVE should be high enough to realize the required coupling strength between the NVE and the flux qubit M;
(iii) in order to implement a quantum memory with high fidelity, the quantum computing (flux qubit C) and memory (NVE) units should be separated by an appropriate distance. 

First, it is possible to couple two flux qubits with a tunable coupling strength $J_{t}$ by using a SQUID or ancilla flux qubit. Experimentally, there are several methods for realizing this tunable coupling. For example, dc-pulse control is a widely used technique for modulating the coupling strength between two flux qubits. By varying the flux piercing the superconducting loop of the ancilla qubit, the coupling strength can be tuned almost from zero to $2\pi\times100$ MHz \cite{14,15,16}.
 
Second, based on recent experiments \cite{12.5}, the coupling
strength between an ensemble of approximately $3\times10^{7}$
NV centers and flux qubit, $g$ can reach up to 35 MHz. Here the value of $g$ is much larger than the decay rate of the flux qubit
($\gamma_{\rm FQ}\sim$ 1 MHz).
With technical advances, the increasing number of the NV centers in each ensemble (or the larger size of the flux qubit) will further enhance the coupling strength. In our proposal, we consider a flux qubit M with size $L_{\rm M}\simeq5$ $\mu$m and a density of NV centers $n>10^{16}$ cm$^{-3}$ \cite{11}. Then, we can estimate the coupling strength between flux qubits and NVE, $g\simeq35$ MHz, which yields a quantum-storage-operation time $T\sim10^{-2}\mu$s. In addition, the dephasing and decay times of the flux qubits made so far are usually order of 1 to $100\,\mu$s. The NV center sample has relatively long decay and dephasing times $T_{\rm 1NV}\sim$ 1 ms and $T_{\rm 2NV}\sim 10^{-2}$ ms. Therefore, in our proposal, the quantum-storage-operation time $T\sim10^{-2}\mu$s is much shorter than the coherence times of flux qubits and NV centers. However, it should also be noticed that the decoherence of the NVE increases when the density of the NV centers increases. Thus, in order to obtain a high-storage-fidelity, it is still required to suppress the decoherence of the NVE. This decoherence is normally induced by the dipole interaction between the redundant nitrogen spins and the NV centers, which is due to the low nitrogen-to-NV conversion rate. Fortunately, this problem could be overcome by applying an external driving field to the electron spins on the redundant nitrogen atoms, which leads to an increased coherence time of the NVE if the nitrogen spins are flipped by the spin-echo pulses on a time scale much faster than the flip-flop processes \cite{11}. 

\begin{figure*}
\includegraphics[width=0.49\textwidth]{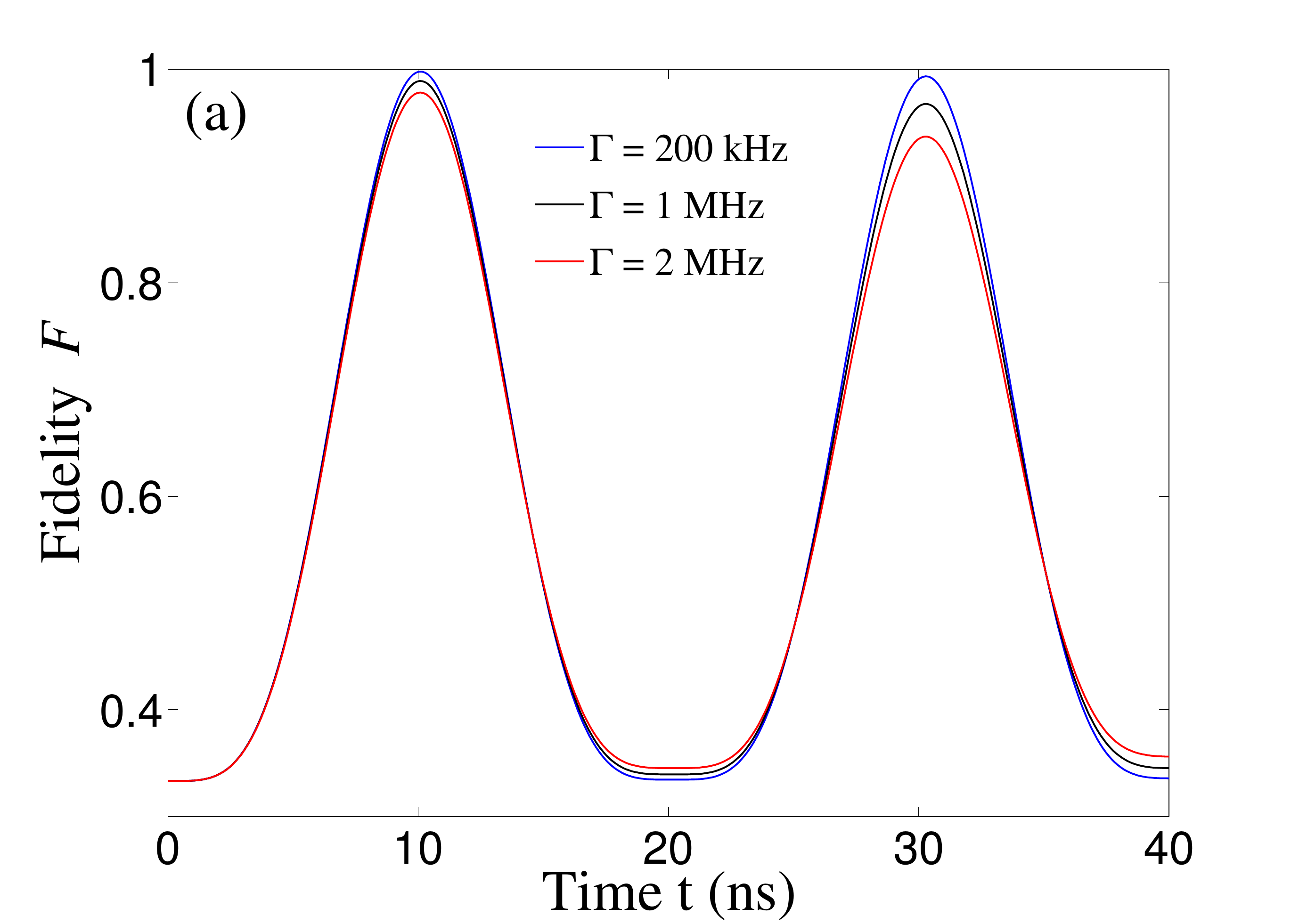}
\includegraphics[width=0.49\textwidth]{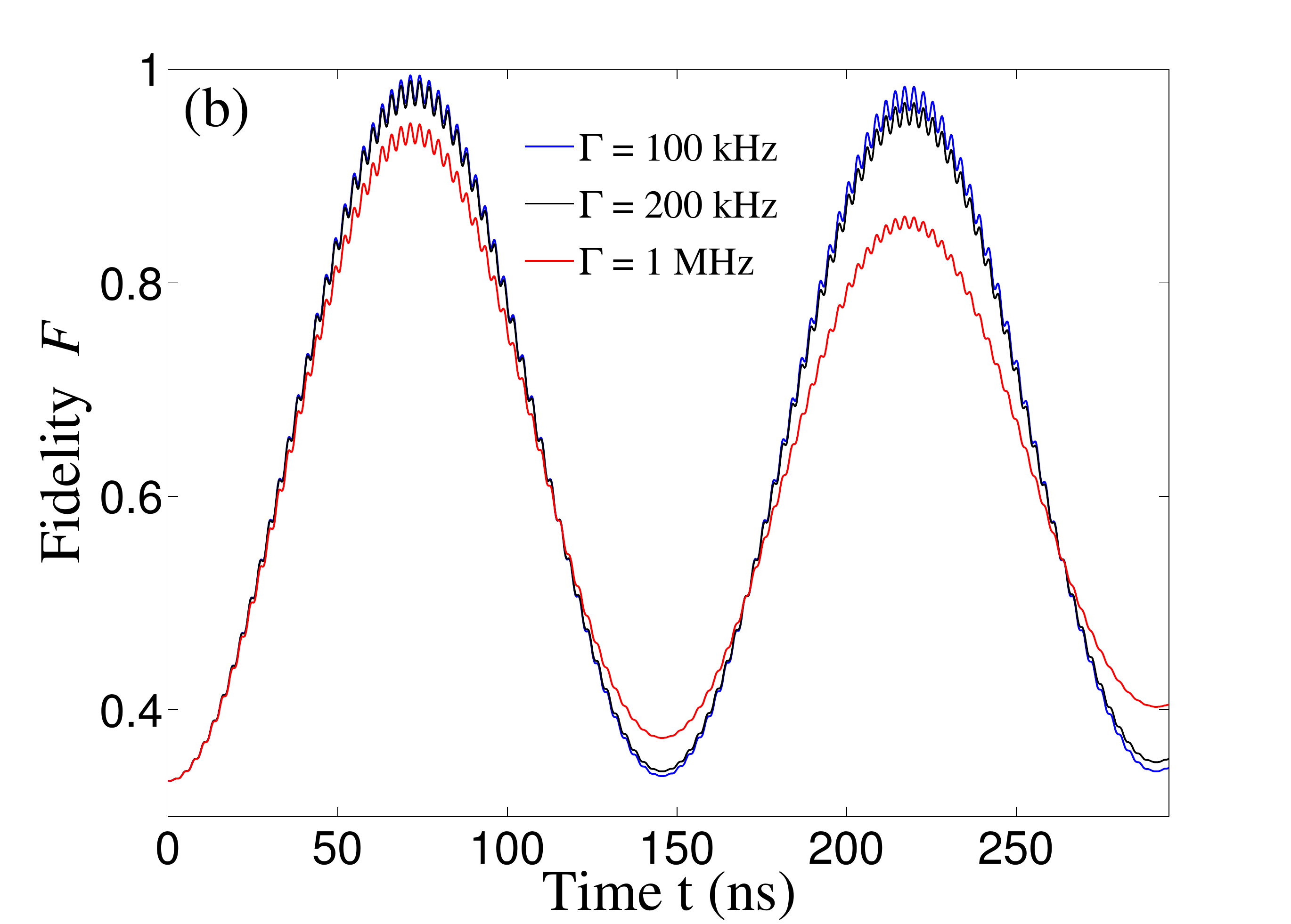}
\caption{(Color online) The fidelities of quantum storage versus time $t$ in the case of resonant (a) and dispersive (b) interactions. The system parameters used here are same as that in Fig.\,\ref{fig7} and the decay rates of flux qubits ${\rm M}$ and ${\rm C}$, $\gamma_{\rm C}=\gamma_{\rm M}=\Gamma$. Note that the decay rate of the NV-center, which has been ignored, is negligible compared with the decay rate of flux qubit.}\label{fig8}
\end{figure*}

To show clearly how the quantum-storage-fidelities are modified when we include decoherence in the system, we now write the full phenomenological quantum Master equation 
\begin{align}
\label{eq15}
\!\!\!\dot{\rho}=-i[H_{\rm tot},\rho]+\sum_{j=M,C}\left[\gamma_{j}\tilde{\sigma}^{-}_{j}\rho\tilde{\sigma}^{\dagger}_{j}-\frac{\gamma_{j}}{2}\left\{\tilde{\sigma}^{\dagger}_{j}\tilde{\sigma}^{-}_{j},\rho\right\}\right],
\end{align}
where $\gamma_{j}$ is the decay rate of flux qubit $j$. Notice that the decay rate of the NV-centers has been ignored because it is much smaller than the decay rate of the flux qubit. In Fig.\,\ref{fig8} we plot the quantum-storage-fidelities $F$ versus time $t$, for different decay rates $\Gamma$ ($\gamma_{M}=\gamma_{c}=\Gamma$). It is shown there that the quantum-storage-fidelities $F$ are robust regarding decoherence of the hybrid system. High-fidelity quantum storage can still be realized even when the decay rate $\Gamma=1$ MHz (200 kHz) in the case of resonant (dispersive) interaction.

Finally, as shown in our numerical calculations (Fig.\,\ref{fig7}), it is necessary to separate the quantum computing (flux qubit) and memory (NVE) units with an appropriate distance ($d_{\rm N}\gtrsim8$ ${\mu}$m), in order to obtain high quantum-storage-fidelity. According to related experiments on coupled-flux qubits \cite{14.5}, the above condition is feasible with current technology. For our case, we can choose an ancilla flux qubit with a length of 5 $\mu$m to connect the flux qubits M and C. Together with the sizes of the flux qubits M and C, the condition $d_{\rm N}>8$ $\mu$m can be satisfied. 
\section{conclusion}
In conclusion, we have proposed a spin-based
quantum memory for a flux qubit based on a hybrid flux qubit and a NV-center system. We have
shown that this proposal can provide high-fidelity quantum storage under realistic conditions, both in the resonant and the dispersive-interaction cases. 
We argue that our proposal can effectively eliminate the mutual influence between
the quantum computing and quantum storage processes by separating the computing and memory units. Thus, high-fidelity quantum storage can be realized in our proposal without needing any additional control pulses on the computing or memory units. Moreover, the quantum computing and memory units can be respectively integrated, which has practical applications in the realization of large-scale quantum memory devices.

\begin{acknowledgements}
XYL thanks S. Ashhab and Jian Ma for valuable discussions. FN was partially supported by the ARO, RIKEN iTHES program, JSPS-RFBR contract No.~12-02-92100, Grant-in-Aid for Scientific Research (S), MEXT Kakenhi on Quantum Cybernetics, and the JSPS via its FIRST program. JQY is partially supported by the National Basic Research Program of China Grant No.~2009CB929302 and the National Natural Science Foundation (NSF) of China Grant No.~91121015. WC was supported by the RIKEN FPR Program. XYL was supported by Japan Society for the Promotion of Science (JSPS) Foreign Postdoctoral
Fellowship No. P12204 and the NSF of China under grant number 11005057.
\end{acknowledgements}

\end{document}